\documentclass[12pt]{article}

\begin{document}

\begin{titlepage}
\begin{flushright}
NSF-ITP-02-59\\
ITEP-TH-38/02\\
\end{flushright}

\begin{center}
{\Large $ $ \\ $ $ \\
Nonspherical Giant Gravitons and Matrix Theory}\\
\bigskip\bigskip\bigskip
{\large Andrei Mikhailov}
\footnote{On leave from the Institute of Theoretical and
Experimental Physics, 117259, Bol. Cheremushkinskaya, 25,
Moscow, Russia.}\\
\bigskip
Kavli Institute for Theoretical Physics,\\
University of California, Santa Barbara, CA 93106\\
\vskip 1cm
E-mail: andrei@kitp.ucsb.edu
\end{center}
\vskip 1cm
\begin{abstract}
We consider the plane wave limit of the nonspherical giant gravitons.
We compute the Poisson brackets of the coordinate functions and find
a nonlinear algebra. We show that this algebra solves the supersymmetry
conditions of the matrix model. This is the generalization of the
algebraic realization of the spherical membrane as the ``fuzzy sphere''.
We describe finite dimensional representations of the algebra 
corresponding to the fuzzy torus.

PACS: 11.25.-w, 11.27.+d, 11.30.Pb
\end{abstract}
\end{titlepage}

\section{Introduction.}
Brane polarization \cite{Myers} provides a remarkable link
between algebra and geometry. The simplest example of this
phenomenon is the realization of the spherical D2 brane
stabilized by the Ramond-Ramond flux as the static configuration
of $N$ D0 branes. The worldvolume theory of the $N$ D0 branes
is the theory of nine $N\times N$ matrices $X_i(t)$ describing the
trajectories of the zero-branes\cite{BoundStates}.
For the branes moving in flat
empty space the equations of motion imply for the static
solutions that the matrices
$X_i(t)$ for $i=1,\ldots,9$ commute with each other. Their
common eigenvalues $x_{i,k}(t)$, $k=1,\ldots, N$ describe
the motion of $N$ $D0$ branes.
But in the curved space with the Ramond-Ramond fluxes there
are static solutions with $[X_i, X_j]\neq 0$. For example,
for the space-time with the constant Ramond-Ramond field
strength there are static configurations with $X_1$, $X_2$, $X_3$
generating the $N$-dimensional representation of the algebra
$su(2)$:
\begin{equation}\label{FuzzySphere}
[X_i,X_j]=r X_k
\end{equation}
These configurations are known as "fuzzy spheres".
In the limit $N\to\infty$ with the fixed $Nr$ such solutions are
interpreted as spherical D2 branes. Therefore D2 branes are solutions
of the worldvolume theory of the D0 branes.
Of course, D0 branes can also be represented as the solutions of the
worldvolume theory of the D2 branes. We can think of them as spherical
D2 branes carrying the topologically nontrivial $U(1)$ bundle.

It turns out that in curved spaces like AdS massless particles
can polarize into massive branes \cite{MGST}. Polarized massless
particles are called "giant gravitons". These giant gravitons
were originally described as continuous branes of the
spherical shape. But massless particles in eleven dimensions can be described
by the Matrix theory \cite{CH,BFSS,Susskind}, therefore we expect that
giant gravitons have an algebraic description.
Recently the authors of \cite{BMN} found the matrix description
of the plane wave geometries which are certain limits of
$AdS_7\times S^4$. It was shown in \cite{BMN} that the giant
gravitons represented by the M2 branes polarized in the direction
of $S^4$ survive in the  plane wave limit and are indeed described
in the corresponding matrix model as  fuzzy spheres (\ref{FuzzySphere}).
The algebraic description of the spherical branes in $AdS_7\times S^4$
and $AdS_4\times S^7$ was found in \cite{JanssenLozano}.

One of the interesting properties of the giant gravitons
is that unlike the polarized D0 branes they are deformable.
We have found in \cite{AM} that spherical giant
gravitons are representatives of the families of non-spherical
solutions parametrized by the holomorphic functions.
In this paper we will argue that the non-spherical giant gravitons
correspond in the matrix model to finite-dimensional representations of the
nonlinear algebra:
\begin{eqnarray}\label{Algebra}
&& [X,\bar{X}]=2X_3\nonumber\\
&& [X_3,X]=X-\bar{\phi}(\bar{X})\\
&& [X_3,\bar{X}]=-\bar{X}+\phi(X)\nonumber
\end{eqnarray}
where $\phi(X)$ is an arbitrary function.
We will start with considering the nonspherical M2 brane preserving
$1\over 4$ of the supersymmetries. We will describe the plane
wave limit of this configuration and compute the Poisson
brackets of the worldsheet coordinates.
We will find that (\ref{Algebra}) is the algebra of the coordinate functions
on the spacial slice of the M2 brane. We then consider
the supersymmetry conditions on the $1\over 4$ BPS configurations
in the matrix theory which were formulated by Dongsu Bak in \cite{Bak}
(see also \cite{SugiyamaYoshida,HyunShin}).
We show that the algebra (\ref{Algebra}) solves these conditions.
As an example we consider the matrix realization of a toroidal giant
graviton. 

\section{Giant gravitons in pp wave background.}
\subsection{Giant gravitons in the Penrose limit.}
To study the non-spherical giant gravitons in $AdS_7\times S^4$
it is convenient to embed $AdS_7\times S^4$ in the flat space with
the signature $(2,11)$.
Consider the flat space ${\bf R}^{2+11}$ with the metric
\begin{equation}
ds^2=\sum\limits_{A=1}^6 dX_A^2-dX_7^2-dX_8^2+
     \sum\limits_{I=1}^5 dY_I^2
\end{equation}
We embed $AdS_7\times S^4$ as the intersection of two
quadrics:
\begin{equation}\label{Embedding}
\begin{array}{r}
X_7^2+X_8^2-\sum\limits_{i=1}^6 X_i^2=4\\
\sum\limits_{i=1}^5 Y_i^2=1
\end{array}
\end{equation}
We will be interested in the Penrose limit of the non-spherical giant
gravitons. We will start with considering the Penrose limit of the embedding
(\ref{Embedding}). It corresponds to the following approximate solution:
\begin{eqnarray}
&& X_7+iX_8=2\exp\left[i\left(x_++{\epsilon^2\over 4}x_-\right)+
{1\over 8}\epsilon^2\sum\limits_{a=1}^6 x_i^2\right]\\[5pt]
&& Y_4+iY_5=\exp\left[2i\left(x_+-{\epsilon^2\over 4}x_-\right)-
{1\over 2}\epsilon^2\sum\limits_{a=1}^3 y_a^2\right]\\[5pt]
&& X_a=\epsilon x_a, \;\;\; a=1,\ldots,6,\;\;\;
   Y_a=\epsilon y_a, \;\;\; a=1,2,3
\end{eqnarray}
The metric is:
\begin{equation}
ds^2=-4dx_+ dx_--4 (dx_+)^2\left(\sum y_a^2 +{1\over 4}\sum x_a^2\right)
+\sum dx_a^2 + \sum dy_a^2
\end{equation}
We will consider the giant gravitons which expand inside $S^4$.
According to the Section 5 of \cite{AM}, the $1\over 4$-BPS giant gravitons
in $AdS_7\times S^4$ correspond to the holomorphic
curves. Given the equation of the curve $F(z_1,z_2)=0$ the equation of
the $M2$ brane worldvolume is given by
\begin{equation}\label{Curve}
F\left({Y_4+iY_5\over (X_7+iX_8)^2},{Y_1+iY_2\over (X_7+iX_8)^2}\right)=0
\end{equation}
In the Penrose limit this reduces to:
\begin{equation}\label{GGPenrose}
{|y|^2+y_3^2\over 2} + ix_- = W(e^{-2ix_+}y)
\end{equation}
where $W$ is a holomorphic function.
It is convenient to consider separately the real and the imaginary parts of
this equation:
\begin{eqnarray}\label{ShapeOfMembrane}
&& |y|^2+y_3^2=W(e^{-2ix_+}y)+\overline{W}(e^{2ix_+}\bar{y}) \\[5pt]
&& x_-={1\over 2i}(W(e^{-2ix_+}y)-\overline{W}(e^{2ix_+}\bar{y}))
\end{eqnarray}
This determines the shape of the membrane worldvolume in terms of a single
holomorphic function $W$.

\subsection{Lightcone gauge and Poisson brackets.}
Lightcone gauge is useful in a situation when there exists a 
lightlike Killing vector field $\xi$, $\xi_{\mu}\xi^{\mu}=0$:
\begin{eqnarray*}
\nabla_{\mu}\xi_{\nu}+\nabla_{\nu}\xi_{\mu}=0\\[5pt]
{\cal L}_{\xi}G_4=0\\[5pt]
\end{eqnarray*}
where $G_4$ is the four form field strength.
The required lightlike Killing vector in the plane wave background
is $\xi={\partial\over\partial x_-}$.
It satisfies the additional constraints which simplify the formalism:
\begin{eqnarray}\label{AdditionalConstraints}
\nabla_{\mu}\xi_{\nu}=0\\[5pt]
\iota_{\xi}G_4=0\label{horizontal}
\end{eqnarray}
The first constraint implies that the one-form $g_{\mu\nu}\xi^{\nu}$
is a total derivative: 
$$g..\left({\partial\over\partial x_-}\right)=-4dx_+$$
 We will choose $x_+$ to be a time coordinate on the worldsheet.
The spacial slices $x_+=\mbox{const}$ of the membrane worldvolume have a
naturally defined symplectic structure\cite{BST,dWHN}.
This allows to formulate the theory of membranes as the
dynamics on the space of functions $X_i$ with the
potential expressed in terms of the Poisson brackets.
Let us  review the construction of the symplectic form.
The momentum corresponding to the Killing vector field 
$\partial\over\partial x_-$ can be computed as an integral
\begin{equation}
P_+=\int\limits_{M2,\;\; x_+=const}\omega
\end{equation}
This defines the two-form $\omega$ which should be taken as the symplectic
form. It is essential that the momentum corresponding to 
$\partial\over\partial x_-$ is carried only by the Born-Infeld term;
because of (\ref{horizontal}) there if no contribution from the 
Wess-Zumino term.
We can describe $\omega$ more explicitly. 
Suppose that we want to compute the integral
of $\omega$ over the small element of the membrane area at $x_+=0$.
We should then take the area of this element and
multiply by the $x_+$-component of its velocity which should be
orthogonal to the surface of the membrane. Suppose that $v_{\perp}$ denotes the vector
on the  membrane worldvolume which is orthogonal to the slice $x_+=const$. Then,
\begin{equation}
\omega={dx_+(v_{\perp})\over |v_{\perp}|}\; \mbox{Area}
\end{equation}
where Area is the two-form which measures the area of the membrane.

Let us compute the symplectic structure for the membrane
described by the Eq. (\ref{GGPenrose}).
We will parametrize the worldvolume of the membrane by
$(x_+,y,\bar{y})$. The metric on the worldvolume is
\begin{equation}
ds^2|_{M2}=-4 (|y|^2+y_3^2)(dx_+)^2 -
{2\over i}(W' d(e^{-2ix_+}y)-\overline{W'}d(e^{2ix_+}\overline{y})) dx_+ +
dy d\overline{y} +(d y_3)^2
\end{equation}
where $y_3=\pm\sqrt{W+\overline{W}-|y|^2}$. One can take
\begin{equation}\label{vPerp}
v_{\perp}={\partial\over\partial x_+}
-2i e^{-2ix_+}W' {\partial\over\partial\bar{y}}
+2i e^{2ix_+}\overline{W'}{\partial\over\partial y}
\end{equation}
with
\begin{equation}
dx_+(v_{\perp})=1,\;\;\;
||v_{\perp}||^2=-4(W+\overline{W}+|W'|^2-yW'-\overline{y}\overline{W'})
\end{equation}
The area element of the slice $x_+=const$ is
\begin{equation}
\mbox{Area}=\sqrt{1+4|\partial y_3/\partial y|^2}dy\wedge d\bar{y}=
{\sqrt{W+\overline{W}+|W'|^2- y W'
-\bar{y}\overline{W'}}\over y_3}dy\wedge d\bar{y}
\end{equation}
This leads to the simple expression for the symplectic form:
\begin{equation}
\omega={dy\wedge d\bar{y}\over y_3}
\end{equation}
The Poisson brackets computed with this symplectic form are:
\begin{equation}\label{PoissonBrackets}
\begin{array}{l}
\{y,\bar{y}\}=2iy_3\\[5pt]
\{y_3,y\}=i(y-\overline{W'}(\bar{y}))\\[5pt]
\{y_3,\bar{y}\}=-i(\bar{y}-W'(y))
\end{array}
\end{equation}
In the next section we will verify that this algebra satisfies the
supersymmetry condition of the Matrix theory.

The time dependence of the worldsheet coordinates can be read from
(\ref{vPerp}):
\begin{equation}\label{TimeDependence}
{dy\over dt}=ie^{it}\overline{W'}(e^{it}\bar{y})
\end{equation}
where we have denoted $t=2x_+$. This corresponds to the supersymmetric
trajectory of the superparticle in the constant magnetic field. Indeed,
let us introduce $z=e^{-it}y$.
We can interpret (\ref{TimeDependence}) as coming from the Lagrangian:
\begin{eqnarray}\label{SuperParticle}
S=\left|{dz\over dt}\right|^2+
i\left(z{d\over dt}\bar{z}-\bar{z}{d\over dt}z\right)+|z|^2+|W'(z)|^2-
zW'(z)-\bar{z}\overline{W'}(\bar{z})-\nonumber\\[5pt]
-iC_{ij}\bar{\xi}^i{d\over dt} \xi^j +C_{ij}\bar{\xi}^i\xi^j-
{1\over 2}W''(z)\varepsilon_{ij}\xi^i\xi^j+
{1\over 2}\overline{W''}(\bar{z})\varepsilon_{ij}\bar{\xi}^i\bar{\xi}^j
\end{eqnarray}
with the supersymmetry transformations:
\begin{equation}
\begin{array}{l}
\delta\xi^i=\left( {dz\over dt}+iz\right)C^{ij}\bar{\epsilon}_j+
i\overline{W'}(\bar{z})\varepsilon^{ij}\epsilon_j\\[5pt]
\delta z=i\epsilon_i\xi^i
\end{array}
\end{equation}
where the supersymmetry parameter $\epsilon_i$ is subject to the
reality condition
\begin{equation}
C^{ij}\bar{\epsilon}_j=-\varepsilon^{ij}\epsilon_j,\;\;\;
C=\left(\begin{array}{cc}1&0\\0&-1\end{array}\right)
\end{equation}
It would be interesting to learn if the Lagrangian (\ref{SuperParticle}) 
has any applications to the matrix description of the giant graviton 
besides giving the supersymmetric trajectories.

\section{Matrix Theory.}
\subsection{Supersymmetry of the massive matrix model.}
Matrix theory describes M theory compactified on a lightlike circle:
\begin{equation}
x_-\sim x_-+2\pi R_-
\end{equation}
The matrix model for the M-theory plane waves was introduced
in \cite{BMN}. In this section we will be interested in the solutions
of this matrix model with $x_1=\ldots=x_6=0$. The bosonic sector
of the model is:
\begin{equation}
\begin{array}{rl}
S=\int\; dt\;\mbox{Tr}&\left[\sum\limits_{j=1}^3
{1\over 2(2R_-)} (D_0 y^j)^2+{(2R_-)\over 4}\sum\limits_{j,k=1}^3
[y^j,y^k]^2
-\right.\\[5pt] &\left.-
{2\over (2R_-)}
\sum\limits_{j=1}^3 (y^j)^2-2i\sum\limits_{j,k,l=1}^3
\epsilon_{jkl}\mbox{Tr}\;y^j y^k y^l\right]
\end{array}
\end{equation}
The generators of the supersymmetry transformation are the
nine-dimensional Majorana spinors $\epsilon$
explicitly depending on time:
\begin{equation}
\epsilon(t)=e^{-{1\over 2}\gamma_{123} t}\epsilon_0
\end{equation}
The condition for $\epsilon$ to annihilate the solution is:
\begin{equation}\label{SUSY}
\left({1\over 2R_-}D_0 y^i\gamma^i +
      {2\over 2R_-} \sum_{i=1}^3 y^i\gamma^i\gamma_{123}
     +{i\over 2}[y^i,y^j]\gamma_{ij}\right)\epsilon(t)=0
\end{equation}
The spherical giant graviton preserves all the supersymmetry of
the matrix model. It corresponds to constant $y^i$ generating a representation
of $su(2)$:
\begin{equation}
[y^i,y^j]={i\over R_-} \epsilon^{ijk} y^k\label{Spherical}
\end{equation}
We are interested in the solutions which respect only half
of the supersymmetry generators. The preserved supersymmetries
are generated by the parameters $\epsilon$ satisfying the condition
\begin{equation}
\gamma^3\epsilon=\epsilon
\end{equation}
For such $\epsilon$ the condition (\ref{SUSY}) is equivalent to:
\begin{eqnarray}\label{MatrixEq}
D_0 y + 2iy - 2R_- i[y^3,y]=0\\
y^3 - {R_-\over 2} [y,\bar{y}]=0\\
D_0 y^3=0
\end{eqnarray}
We will choose the gauge $A=0$, $D_0={\partial\over\partial t}$.
The solution to (\ref{MatrixEq}) is
\begin{eqnarray}\label{SolutionToMatrixEq}
&& y^3(t)=y^3(0)={R_-\over 2}[y(0),\overline{y(0)}]\label{yyBar}\\
&& y(t)=e^{-2it} e^{2iR_-ty^3} y(0) e^{-2iR_-ty^3}
\end{eqnarray}
Supersymmetric solutions to the matrix model equations of motion
are subject to the constraint
\begin{equation}
{1\over 2}([\partial_t y, \bar{y}]+[\partial_t \bar{y},y])
+[\partial_t y^3, y^3]=0
\end{equation}
The solution (\ref{SolutionToMatrixEq}) satisfies this constraint
if and only if
\begin{equation}\label{FromConstraint}
[[y^3,y],\bar{y}]={1\over R_-}[y,\bar{y}]
\end{equation}
Equations (\ref{yyBar}) and (\ref{FromConstraint}) were derived by Dongsu
Bak in \cite{Bak}. They determine supersymmetric solutions of the matrix
model which preserve one half of the supersymmetry.
These equations have a solution:
\begin{equation}\label{CommutatorAlgebra}
\begin{array}{l}
[y,\bar{y}]={2\over R_-} y^3\\[5pt]
[y^3,y]={1\over R_-}(y-\overline{W'}(\overline{y}))\\[5pt]
[y^3,\bar{y}]=-{1\over R_-}(\bar{y}-W'(y))
\end{array}
\end{equation}
parametrized by a holomorphic function $W(y)$. This relations determine the
associative algebra corresponding (in Matrix theory) to the Poisson
brackets (\ref{PoissonBrackets}). Therefore it represents the nonspherical
giant graviton.

\subsection{Representations for the toroidal giant graviton.}
It would be interesting to study the representations of the algebra
(\ref{CommutatorAlgebra}).
Our giant gravitons are compact manifolds, which suggests that one should
look for the finite-dimensional representations. For the giant gravitons
with the shape close to spherical, it should be possible to realize the algebra
(\ref{CommutatorAlgebra}) on the space of a finite dimensional representation
of $su(2)$. Indeed, one can always find three
functions $S_1,S_2,S_3$ with the Poisson brackets
$\{S_i,S_j\}=\epsilon_{ijk}S_k$ and the constraint $\sum S_i^2=R^2$ where
$R$ is related to the area of the sphere. (There is no difference between
the round sphere and the deformed sphere, from the point of view of the
symplectic geometry.) The coordinates $y_1$, $y_2$, $y_3$ can be expressed
in terms of $S_1,S_2,S_3$, which means that they act on the
representation space of $su(2)$.

Here we want to consider the example of the toroidal membrane
corresponding to $W(y)=A\log y+C$, with real positive $A$ and $C$.
The equation for the membrane worldvolume follows from
(\ref{ShapeOfMembrane}):
\begin{eqnarray}\label{ToroidalShape}
&& y_3=\pm\sqrt{2C-|y|^2+A\log|y|^2}\\[5pt]
&& x_-=A(-2x_++\mbox{Arg}(y))
\end{eqnarray}
For this solution to make sense after compactification of $x_-$
we have to impose the periodicity condition
\begin{equation}\label{IntegralityForA}
A=2\pi k R_-,\;\;\;\; k\in {\bf Z}
\end{equation}
For large $C$ the solution
(\ref{ToroidalShape}) can be interpreted as 
the spherical $D2$ brane with $k$ fundamental strings
stretching from the north to the south pole.

The corresponding algebra is:
\begin{equation}\label{Toroidal}
\begin{array}{l}
[y,\bar{y}]={2\over R_-}y_3\\[5pt]
[y_3,y]={1\over R_-}(y-A\bar{y}^{-1})\\[5pt]
[y_3,\bar{y}]=-{1\over R_-}(\bar{y}-A y^{-1})
\end{array}
\end{equation}
We will try the following ansatz for the finite 
dimensional representations:
\begin{equation}
y=\sum\limits_{j=1}^N q_j E_{j+1,j}
\end{equation}
where $E_{i,j}$ is the matrix $||E_{i,j}||_{mn}=\delta_{im}\delta_{jn}$
and we put $E_{N+1,N}=E_{1,N}$. For example for $N=4$:
\begin{equation}
y=\left(\begin{array}{cccc}
0   & q_1 & 0   & 0   \\
0   & 0   & q_2 & 0   \\
0   & 0   & 0   & q_3 \\
q_4 & 0   & 0   & 0  \end{array}\right)
\end{equation}
The operators $y,\bar{y}$ and $y_3={R_-\over 2}[y,\bar{y}]$ satisfy
(\ref{Toroidal}) if
\begin{equation}
2|q_j|^2-|q_{j-1}|^2-|q_{j+1}|^2={2\over R_-^2}
\left(1-{A\over |q_j|^2}\right)
\end{equation}
We will consider this equation in the continuous approximation which is
possible when $A>>1$. We denote $\tau_j={j\over R_-\sqrt{A/2}}$ 
and substitute
$|q_j|^2=A(1+f(\tau_j))$. We will get:
\begin{equation}
f''(\tau)=-{f\over 1+f}
\end{equation}
This has a solution
\begin{equation}
\tau=\pm\int{df\over\sqrt{2(E-f+\log(1+f))}}
\end{equation}
The "potential energy" $U(f)=f-\log(1+f)$ is a convex function of $f$ with
the minimum $U(0)=0$. This means that for any $E>0$ $|q_j|^2$ as a function of $j$
will be oscillating around $A$. For finite $A$, the correct boundary conditions at
the turning points will restrict $E$ to belong to a discreet set.
This solution describes the toroidal giant graviton.

Finite dimensional representations of the algebra (\ref{Toroidal})
solve the Matrix theory equations of motion for arbitrary $A$.
But we expect that only the algebras with the integer $A$
specified by the equation (\ref{IntegralityForA}) 
have a physical meaning. Such integrality conditions are necessary 
when the worldsheet is not simply connected. It would be interesting
to understand these conditions from the purely algebraic point of view
--- what is special about the algebra (\ref{Toroidal}) when
$A$ satisfies (\ref{IntegralityForA}) with integer $k$?

Another interesting problem is
 to describe finite dimensional representations for
general $W$. A similar problem was solved in \cite{CornalbaTaylor} for holomorphic
membranes in flat space which lead to infinite-dimensional
representations.

\section{Acknowledgements}
This work was supported in part by the National Science Foundation grant
No. PHY99-07949, RFFI grant No. 00-02-116477 and the
Russian grant for the support of the scientific schools No. 00-15-96557.


\begin{thebibliography}{10}
\bibitem{Myers}{R.C.~Myers, "Dielectric-Branes",
JHEP 9912 (1999) 022, hep-th/9910053.}
\bibitem{BoundStates}{E.~Witten, "Bound States Of Strings
And $p$-Branes", Nucl. Phys. {\bf B460} (1996) 335-350,
hep-th/9510135.}
\bibitem{MGST}{J.~McGreevy, L.~Susskind, N.~Toumbas,
"Invasion of the Giant Gravitons from Anti-de Sitter Space",
JHEP 0006 (2000) 008.}
\bibitem{CH}{M.~Claudson and M.~B.~Halpern, "Supersymmetric Ground State
Wave Functions", Nucl. Phys. {\bf B250} (1985) 689-715.}
\bibitem{BFSS}{T. Banks, W. Fischler, S.H. Shenker, L. Susskind,
"M Theory As A Matrix Model: A Conjecture", 
Phys.Rev. D55 (1997) 5112-5128, hep-th/9610043.}
\bibitem{Susskind}{L.~Susskind, "Another Conjecture about M(atrix)
Theory", hep-th/9704080.}
\bibitem{BMN}{D.~Berenstein, J.~Maldacena and H.~Nastase,
"Strings in Flat Space and PP-Waves from ${\cal N}=4$ Super
Yang-Mills", JHEP {\bf 0204} (2002) 013, hep-th/0202021.}
\bibitem{JanssenLozano}{ B.~Janssen, Y.~Lozano, 
"A Microscopical Description of Giant Gravitons",
hep-th/0207199.}
\bibitem{AM}{A.~Mikhailov, "Giant Gravitons from
Holomorphic Surfaces", hep-th/0010206.}
\bibitem{Bak}{Dongsu Bak, "Supersymmetric Branes in PP Wave
Background", hep-th/0204033.}
\bibitem{SugiyamaYoshida}{K.~Sugiyama and K.~Yoshida,
"BPS Conditions of
Supermembrane on the PP Wave", hep-th/0206132; 
"Giant Graviton and Quantum Stability in Matrix Model
on PP-wave Background", hep-th/0207190.}
\bibitem{HyunShin}{Seungjoon Hyun and Hyeonjoon Shin,
"Branes From
Matrix Theory in PP-Wave Background", hep-th/0206090.}
\bibitem{BST}{E.~Bergshoeff, E.~Sezgin and P.K.~Townsend,
Phys.Lett. {\bf B189} (1987) 75; Ann. Phys. (NY) 185 (1988) 330.}
\bibitem{dWHN}{B.~de~Wit, J.~Hoppe and H.~Nicolai, Nucl. Phys.
 {\bf B 305} (1988) 545.}
\bibitem{CornalbaTaylor}{L.~Cornalba, W.~Taylor, "Holomorphic curves
from matrices", Nucl.Phys. {\bf B 536} (1998) 513-552.}
\end{thebibliography}
\end{document}